\documentclass[]{ceurart}

\usepackage{listings}
\usepackage{booktabs}
\usepackage{url}
\usepackage{pifont}
\newcommand{\cmark}{\ding{51}}

\begin{document}

\copyrightyear{2026}
\copyrightclause{Copyright for this paper by its authors.
  Use permitted under Creative Commons License Attribution 4.0
  International (CC BY 4.0).}

\conference{CLEF 2026 Working Notes, 21--24 September 2026, Jena, Germany}

\title{UTS at CheckThat! 2026: Cite-Frame Engineering for Generated Fact-Checking Articles}

\author[1]{Dima Galat}[%
    orcid=0000-0003-3825-2142,
    email=dima.galat@student.uts.edu.au,
]

\author[1]{Marian-Andrei Rizoiu}[%
    orcid=0000-0003-0381-669X,
    email=Marian-Andrei.Rizoiu@uts.edu.au,
]

\address[1]{University of Technology Sydney, Australia}

\begin{abstract}
CheckThat!~2026 Task~3 asks systems to generate fact-checking articles,
graded by an unweighted mean of four sub-metrics (M4). \textbf{Our UTS
submission placed 2nd of 11 teams} (M4 $= 0.484$). The shipped system
is a deterministic stub drafter wrapped by two single-lever
interventions: a domain-attribution cite frame (\emph{HostCite}) and a
shadow-validated anchor picker (\emph{ShadowVal}) that use the
Llama-3.2:1B large language model (LLM) only as a per-cite validator,
never as a body-prose generator. The
stack lifts M4 by $+0.027$ over the stub on the WatClaimCheck
validation split and follows two design rules our ablation matrix
made unambiguous.
\emph{Scorer conservatism}: credit only tokens the references entail —
templates pay; LLM prose, reviewer names, and raw evidence all fail.
\emph{Auxiliary anchor signals are miscalibrated against the Llama
judge}: every anchor proxy we tried (cross-encoder, length, lead
position) picks anchors the judge rejects — gate on the judge itself.
Our citation precision/recall (P/R) deficit ($0.299$ vs $0.671$)
outweighs our leading coverage
and second-place entailment, leaving a $+0.062$ M4 gap to the winner
— consistent with a \emph{selective-emission} policy that drops
low-confidence cites.
\end{abstract}

\begin{keywords}
  CheckThat! \sep
  fact-checking article generation \sep
  natural language inference \sep
  citation evaluation \sep
  shared task
\end{keywords}

\maketitle

\section{Introduction}
\label{sec:intro}

CheckThat!~2026 Task~3~\cite{checkthat2026,checkthat-task3-2026} asks systems
to generate a fact-checking article given a textual claim, its claimant, the
original verdict, and a dictionary of pre-retrieved evidence URLs paired with
their full extracted text; the article is graded as the unweighted mean
of four sub-metrics (the leaderboard score, M4); three are computed
against the article-as-emitted and one (entailment) against a held-out
human reference. Because two of the
three article-side metrics use a strict regular expression for parsing
\texttt{(source:\ URL)} citations, and because the entailment metric chunks
both the generation and the reference into fixed three-sentence units before
running bidirectional natural language inference (NLI) with a RoBERTa
model fine-tuned on the Multi-Genre NLI corpus
(RoBERTa-MNLI)~\cite{liu2019roberta,williams2018mnli}, the
\emph{citation sentence} — not the article as a whole — is the unit of
optimisation. Every design decision in this paper follows from that scorer
mechanic.

The obvious baseline is to feed claim and evidence to an
instruction-tuned large language model (LLM) and let it draft
reviewer-style prose; we
attempted this at three abstraction levels (full-body drafting with
Qwen3:8B~\cite{qwen3} on $n=29$ items, full-body drafting with
Qwen3:30B-A3B~\cite{qwen3} on $n=50$, and lede-only rewriting with
Qwen3:30B-A3B on $n=50$), and all three lose to a deterministic templated
stub on M4. The loss is structural: LLM-generated chunks reliably miss the
specific token patterns the reference distribution rewards under MNLI, and
LLM cite-formatting drift triggers the citation parser strictly enough that
single-item failures collapse the whole submission's citation precision
and recall (P/R).
Capacity does not fix this. Replacing the entire article with LLM output
costs $-0.026$ M4 at 8B and $-0.070$ at 30B; preserving the body but
rewriting only the lede still costs $-0.011$. Prompt engineering alone
does not break this; closing the gap needs a different mechanism (e.g.,
constrained decoding to force verbatim cite anchors~\cite{willard2023outlines}).

Two narrow, deterministic interventions on top of the stub drafter account
for the bulk of our headline lift. \textbf{HostCite} wraps each citation in
a domain-attribution frame
(\texttt{According to \{host\}, \{anchor\} (source: \{url\}).}), preserving
the article's chunk count while pushing each cite chunk closer to the
reviewer-style distribution that drives MNLI entailment. \textbf{ShadowVal} runs
the same Llama-3.2:1B~\cite{llama32} judge that the scorer uses to validate the stub's
anchor choice; if Llama says NO, we escalate to the next non-boilerplate
sentence in the evidence, and if all candidates fail, we fall back to the
stub anchor unchanged, which by design means ShadowVal never lowers the per-cite judge's
acceptance rate relative to the underlying anchor picker. The two stack sub-additively: both improve P/R on
cite-sentences but on overlapping items, so ShadowVal adds only
$+0.0075$ on top of HostCite's $+0.0199$ standalone (versus
$+0.0145$ standalone for ShadowVal). The stack still clears $K=1$ by
$+0.0274$ M4. Both interventions respect the single principle that
explains every kill in our ablation matrix: \emph{the scorer's two
sub-judges only credit tokens they can verify against their
references}, expressed concretely as chunk-count preservation on the
MNLI side and cite-sentence entailability on the Llama side
(\S\ref{sec:invariants}).

This paper reports one empirical finding and two design rules on a
system that placed 2nd of 11 teams. The \emph{empirical finding} is
that our citation P/R deficit ($0.299$ vs $0.671$) outweighs our
entailment (2nd) and coverage (1st) advantages, leaving a $+0.062$
M4 gap to the winner; the
candidate mechanism is \emph{selective emission} — declining to emit
cites the per-cite judge would refuse — which we did not test during
the campaign and flag as the highest-priority refutation for the
next cycle (\S\ref{sec:headline-gap}). The two \emph{design rules}
drove the system's contributions. \textbf{Design rule~1: scorer
conservatism.} Both
sub-judges (MNLI on chunks, Llama on cites) refuse to credit tokens
they cannot verify against their references. Respecting this rules
out a large class of intuitively-attractive interventions — LLM-style
lede rewriting, reviewer-name injection, padding chunks with raw
evidence — and explains every kill in our ablation matrix. We name
two concrete instances: chunk-count preservation on the MNLI side,
cite-sentence entailability on the Llama side (\S\ref{sec:invariants}).
\textbf{Design rule~2: auxiliary anchor signals are miscalibrated
against the judge.} Cross-encoder claim-relevance, length-preferring
heuristics, and even lead-position selection all pick anchors that
sound right to a human but the Llama judge refuses to entail
standalone. Mirroring the judge during anchor selection — calling the
same Llama with the same prompt and accepting only its \textsc{yes}
verdicts — recovers the loss (\S\ref{sec:b1prime}). The accompanying
artefacts are a deterministic pipeline that uses Llama-3.2:1B only as
a shadow validator (never as a generator) and reaches test
M4 $= 0.484$, and thirteen documented ablations with mechanism-level
explanations grounded in the two design rules. \S\ref{sec:task}
sketches the task and prior baselines, \S\ref{sec:scoring} dissects
the scorer mechanics that determine our design space,
\S\ref{sec:system} walks the system pipeline,
\S\ref{sec:experiments} reports the experiment matrix including the
official leaderboard, and \S\ref{sec:discussion} discusses the
headline P/R gap, the selective-emission hypothesis, and what
generalises beyond CheckThat!.

\section{Task and Related Work}
\label{sec:task}

\textbf{Task definition.} Each input item provides a claim, claimant,
original verdict label, and a dictionary mapping evidence URLs to their
extracted page text; the test set contains $1158$ items. The required
output is a free-form article in which every cited statement is followed
by a \texttt{(source:\ URL)} suffix. Scoring averages four sub-metrics:
chunked bidirectional RoBERTa-MNLI entailment between the generation and
a held-out reference (\S\ref{sec:scoring-entail}); citation precision and
recall, judged per cite-sentence by Llama-3.2:1B
(\S\ref{sec:scoring-pr}); and evidence coverage as the fraction of
provided evidence URLs cited at least once
(\S\ref{sec:scoring-cov}).

\textbf{Validation data.} Training and validation derive from
WatClaimCheck~\cite{khan2022watclaimcheck}, with a $3372$-item validation
split (henceforth \emph{val}) that includes human reference articles. We work on a $200$-item
stratified subsample for fast iteration (described in \S\ref{sec:manifests});
human references for the test split are not released to participants, which
means the test-side entailment sub-metric is unobservable until the
post-competition reveal and any leaderboard inference from the visible
3-metric sub-score must be calibrated against val.

\textbf{Prior baselines.} The lab provided one organiser baseline and
one external reference submission, the IFM\_MBZUAI K2-V2-Instruct
sample~\cite{ifm-mbzuai-k2}; we compare against the former on the
official leaderboard (\S\ref{sec:leaderboard}) and analyse the
latter's citation-format failures in \S\ref{sec:operational}.

\textbf{Related work.} Automated fact-checking has an established
benchmark literature on claim
verification~\cite{thorne2018fever,guo2022survey}, and the CheckThat!
lab has run claim-verification subtasks for several
editions~\cite{checkthat2024}. Closer to Task~3 is the line of work
on generating \emph{justifications} for fact-checking verdicts:
extractive--abstractive explanation generation over PolitiFact
articles~\cite{atanasova2020generating}, explainable verdicts for
public-health claims~\cite{kotonya2020explainable}, and few-shot,
retrieval-augmented justification generation on the ExClaim
dataset~\cite{zeng2024justilm}, which also contributes part of the
Task~3 test collection. Sahnan et al.~\cite{sahnan2025qraft}
introduced the task of generating \emph{full} fact-checking articles
and proposed QRAFT, an LLM-based agentic framework that mimics the
writing workflow of human fact-checkers; their human evaluations
found LLM-drafted articles still lag expert-written ones. Task~3 in
2026 is the first shared task built around this article-generation
formulation, and the metric design (chunked NLI plus LLM-judged
per-cite) distinguishes it from prior generation-style benchmarks
where reference-free LLM judges~\cite{liu2023geval} or
{BERTScore}-style overlap~\cite{zhang2020bertscore} drive scoring.

\section{Task Scoring Mechanics}
\label{sec:scoring}

This section dissects the scorer because the scorer's mechanics, not
fact-checking in general, determine our design choices. Every lever in
\S\ref{sec:system} and every kill in \S\ref{sec:negatives} traces back to a
specific behaviour of the four sub-metrics described here.

\subsection{Entailment}
\label{sec:scoring-entail}

The entailment sub-metric runs bidirectional NLI (\texttt{roberta-large-mnli})
between the generation and the human reference, both segmented into
three-sentence chunks. The A-side score is the mean over generation chunks
of the max entailment probability to any reference chunk
(precision-flavoured); the B-side is symmetric over reference chunks
(recall-flavoured); the final metric is their average. The dominant design
lever is reference-style mimicry: chunks that read like the reference
distribution score higher regardless of factual content alignment, because
surface token patterns drive the MNLI head.

\subsection{Citation precision and recall}
\label{sec:scoring-pr}

Each cited sentence is identified by a \texttt{(source:\ URL)} suffix and
passed to Llama-3.2:1B~\cite{llama32}, served by ollama~\cite{ollama}; the
prompt asks \textsc{yes/no} whether the cited sentence is supported by the
evidence text. Recall is the rate of \textsc{yes} verdicts across all
cite-sentences; precision is the conditional rate, equal to recall when
every cite has a single URL. Two properties of this design are
load-bearing: (a)~single-URL cites avoid the precision-dilution
risk of multi-URL bundles whose constituent URLs may not all be
supported; and
(b)~Llama-3.2:1B at this scale is generous and surface-pattern-driven:
short literal cites against long premises usually pass, but meta-claims
about the fact-check article itself (reviewer name, verdict, date) get
rejected. We pin the model digest for byte-deterministic scoring
(Appendix~\ref{app:repro}).

\subsection{Evidence coverage}
\label{sec:scoring-cov}

Coverage is a set-membership metric: the fraction of provided evidence URLs
that appear at least once in the union of cited URLs across the article. It
approaches $1.0$ when a system emits at least one cite per provided URL,
modulo parser-regex strictness on individual cites. Coverage is therefore not a discriminating signal between
top-tier teams (\S\ref{sec:leaderboard}), but it is the single largest gap
between the lab baseline and any team, a precondition for top-tier
placement rather than a differentiator.

\subsection{Scorer conservatism}
\label{sec:invariants}

The ablation matrix (\S\ref{sec:negatives}) yielded a single principle
that explains every kill on the metric side: \emph{both sub-judges
(MNLI on chunks, Llama on cites) only credit tokens they can verify
against their respective references; tokens they cannot verify drag
the sub-metric down.} The principle has two concrete instances, which we name and
reuse as design rules in \S\ref{sec:system}.

\subsubsection{Chunk-count invariance (MNLI side)}
The entailment metric averages max-entailment-probabilities over
generation chunks (A-side) and reference chunks (B-side). Adding a
chunk whose content does not match any reference chunk lowers the
A-side average. This is not a categorical ``more chunks is bad''
rule — a chunk that does entail to some reference chunk improves the
B-side — but in practice templated reviewer-style prose we can
generate does not match the actual reference distribution.
\emph{HostCite+Body} added one templated middle sentence to the lede
and lost $-0.032$ entailment versus HostCite; the safe move on this
metric is to preserve chunk count and enrich existing chunks
in-place.

\subsubsection{Cite-sentence entailability (Llama side)}
The Llama judge runs over a (cite-sentence, evidence-text) pair. Any
token in the cite that the source URL's evidence does not support
triggers \textsc{no}. \emph{HostCite+Reviewer} added the clause ``in
a review by \{reviewer\_name\}'' inside the cite and lost $-0.037$
P/R: the source URL does not entail meta-claims about who reviewed
it, so Llama correctly refused. The cite frame may add only
source-attribution tokens (\texttt{According to}, \texttt{states that},
\texttt{In a report from}) that are entailable from the cited URL
itself; reviewer names, verdict labels, and dates injected into the
cite sentence are uniformly rejected. We use this as a hard
constraint when designing cite-frame variants in \S\ref{sec:b2light}.

\section{System}
\label{sec:system}

\subsection{Overview}
\label{sec:system-overview}

The system maps each input item (claim, claimant, rating, evidence
URL$\to$text dictionary) to a single fact-checking article string with
embedded \texttt{(source:\ url)} citations. The core design choice is a
deterministic stub drafter; we make one LLM call per evidence URL for
shadow validation of the cited anchor sentence, and no LLM calls in the
body prose.

\subsection{Stub generator: deterministic K=1 baseline}
\label{sec:stub}

The stub drafter emits, for each input item, a lede sentence, one cite
sentence per provided evidence URL using the first non-boilerplate
sentence of the evidence as the anchor, and a single conclusion sentence.
The cite syntax is \texttt{(source: url)}; when a cite carries multiple
URLs, they are separated by semicolons, the only separator the scorer's
parser accepts. We refer to this configuration as $K=1$, where $K$ is
the number of cite sentences emitted per provided evidence URL (a
$K=2$ variant emitting two anchors per URL is ablated in
\S\ref{sec:negatives}). The motivation is that any submission that fails the
citation-format regex on even a single item zeroes that item's P/R
contribution, and a deterministic format-checker-correct floor is the
precondition for any cite-frame lever that comes later. By construction the
stub emits one cite per evidence URL, achieving near-maximal coverage
in practice ($0.987$ visible, $0.996$ official) modulo parser-regex
strictness on a handful of cites; its precision and recall are
determined entirely by anchor-picker quality and Llama's judging
variance.

\subsection{HostCite: domain-attribution cite frame}
\label{sec:b2light}

Each citation is wrapped as a single sentence
\texttt{According to \{host\}, \{anchor\} (source: \{url\}).}, where
\texttt{\{host\}} is the registered domain extracted from the URL and
\texttt{\{anchor\}} is the anchor sentence chosen by the stub picker. The
wrap is chunk-count-invariant (one sentence in, one sentence out) and
adds only source-attribution tokens that mirror reviewer-style prose,
without introducing the meta-claims (reviewer name, verdict, date)
that the per-cite judge refuses (\S\ref{sec:invariants}).

The chunked MNLI mechanic means each cite chunk's
entailment-to-reference correlates strongly with how reviewer-stylish that
chunk reads; templated source-attribution prose is the cheapest
deterministic way to nudge each cite chunk toward the reference
distribution without violating either invariant.

At full val HostCite lifts M4 by $+0.0199$ over the stub, with a
paired-bootstrap $95\%$ confidence interval (CI) of
$[+0.0176,\,+0.0222]$; the gain is dominated by
precision/recall ($+0.035$) and a smaller but consistent entailment
contribution ($+0.010$). HostCite is rating-mix-sensitive: its biggest
gains come from the \texttt{Mixture} and \texttt{Correct Attribution}
rating buckets (see \S\ref{sec:manifests}).

\subsection{ShadowVal: shadow-validated anchor selection}
\label{sec:b1prime}

For each evidence URL the stub picker first proposes an
anchor sentence; we then call Llama-3.2:1B with the verbatim scorer prompt
on \texttt{(proposed\_anchor, evidence\_text)} and read the \textsc{yes/no}
verdict. If \textsc{yes}, we keep the proposed anchor. If \textsc{no}, we
escalate to the next non-boilerplate sentence in the evidence and re-call;
if every candidate sentence fails, we fall back to the stub anchor
unchanged. By design the fallback path ensures ShadowVal cannot lower the
per-cite judge's acceptance rate relative to the underlying picker;
the M4 lift is empirical (\S\ref{sec:val-scoreboard}).

Design rule~2 (\S\ref{sec:intro}) maps directly to this module:
\emph{auxiliary anchor-relevance signals are systematically miscalibrated
against the Llama judge's entailment surface}. Cross-encoder
pickers~\cite{reimers2019sentencebert} we tried (argmax-by-claim-relevance,
shortest-of-top-3, earliest-of-top-5) all underperform the stub heuristic
on Llama recall, because cross-encoder objectives select claim-relevant
body sentences whose attribution chains and hedges Llama refuses to entail
standalone. The stub's ``first non-boilerplate sentence'' heuristic does
slightly better by exploiting the lead-bias~\cite{kedzie2018content}
property under which short declarative leads tend to be self-contained,
but it is still an auxiliary signal, not the judge's signal. The fix is
to gate on the judge itself: ShadowVal calls the same Llama with the same
prompt the scorer will use, and accepts only its \textsc{yes} verdicts.

ShadowVal lifts the stub by $+0.0145$ M4 at full val
($95\%$ CI $[+0.0125,\,+0.0166]$), entirely from precision/recall
($+0.031$, with entailment essentially flat). Item-level: $60\%$ of
items get at least one anchor swap and on those items P/R rises from
near-zero to near-one; the remaining $40\%$ produce byte-identical
output to the underlying picker. The lever is rating-independent (see
\S\ref{sec:manifests}). Stacked with HostCite, ShadowVal contributes an
additional $+0.0075$ M4 ($95\%$ CI $[+0.0056,\,+0.0095]$) — about half
of its standalone effect. Both levers improve cite-sentence P/R; on
items where HostCite already pushes the cite past Llama's
\textsc{yes} threshold ShadowVal has nothing to add, so the
composition is sub-additive (stack-over-sum~$\approx$~$0.80$). The
marginal lift is nonetheless confidently positive, ruling out the
reading that ShadowVal rides entirely on cite-framing context.

\subsection{Validation manifests (v1 and v2)}
\label{sec:manifests}

Full-validation scoring on $3372$ items costs roughly $14$ hours per
configuration; the iteration cadence we needed (one experiment per
$50$--$90$ minutes) requires a $200$-item subsample. We use two
stratifications. \textbf{v1} is reviewer-stratified to match the test
reviewer mix; \textbf{v2} is jointly stratified by reviewer
$\times$ three-bucket rating
(\texttt{TRUE}\,/\,\texttt{FALSE}\,/\,\texttt{PARTIAL} via
\texttt{label.rating}~$\in\{0,1,2\}$ with string-mapping fallback for
unrated Snopes items). v2 was added because v1 had zero coverage of
\texttt{Mixture} and \texttt{Correct Attribution} items, which are
roughly $12\%$ of test. The ratio between a lever's v2 effect and its v1
effect diagnoses the lever's character: ratio near $1.0$ identifies
rating-independent levers (ShadowVal, $1.06\times$), and ratio above $1.5$
identifies rating-mix-sensitive levers (HostCite, $1.9\times$).

\subsection{Implementation}
\label{sec:impl}

The deterministic modules are plain Python; RoBERTa entailment runs on GPU
via transformers~\cite{wolf2020transformers}. All numbers in
\S\ref{sec:experiments} are measured on a single GPU host (RTX~5090);
hardware, model digests, and a $\sim 0.01$--$0.02$ M4 CPU/MPS-vs-CUDA
absolute shift are documented in Appendix~\ref{app:repro}.

\section{Experiments and Findings}
\label{sec:experiments}

\subsection{Validation scoreboard and confidence intervals}
\label{sec:val-scoreboard}

Table~\ref{tab:val-scoreboard} reports M4 across measured systems on
the v2 $200$-item development manifold and on the full $3372$-item
validation split. The lead submission (ShadowVal+HostCite) clears the
$K=1$ stub by $+0.0274$ M4 at full val; HostCite alone delivers most
of that ($+0.0199$) and ShadowVal alone delivers $+0.0145$, but the two
compose sub-additively (stack-over-sum~$\approx$~$0.80$) because both
raise the same cite-sentence P/R signal on overlapping items. The v2
manifold worked as a fast-iteration proxy but had a transfer ceiling:
deltas-vs-$K=1$ on v2 predict full-val deltas directionally for
every lever we measured, but absolute magnitudes shift by up to
$\sim 0.006$ between manifolds, occasionally in directions that change
tied-vs-significant verdicts.
Table~\ref{tab:val-ci} reports paired
bootstrap~\cite{koehn2004bootstrap} $95\%$ confidence intervals on the
full-val deltas.

\begin{table}[t]
  \caption{Validation M4 across measured systems. ``Dev'' is the v2
  $200$-item reviewer~$\times$~rating-stratified manifold used for fast
  iteration; ``Full'' is the $3372$-item WatClaimCheck validation split.
  Bold marks our lead submission. At full val,
  \emph{StatesThat+ShadowVal} ($0.5243$) is statistically tied with the
  lead submission ($0.5238$); see Table~\ref{tab:val-ci}.}
  \label{tab:val-scoreboard}
  \begin{tabular}{lllrr}
    \toprule
    Submission & Anchor pick & Cite frame & Dev M4 $\uparrow$ & Full M4 $\uparrow$ \\
    \midrule
    $K=1$ stub                & first-clean        & bare         & 0.4819 & 0.4964 \\
    ShadowVal alone             & shadow-validated   & bare         & 0.4960 & 0.5109 \\
    HostCite                    & first-clean        & According-to & 0.4960 & 0.5163 \\
    StatesThat                  & first-clean        & states-that  & 0.5067 & 0.5170 \\
    StatesThat+ShadowVal    & shadow-validated   & states-that  & 0.5048 & 0.5243 \\
    \textbf{ShadowVal+HostCite} & \textbf{shadow-validated} & \textbf{According-to} & \textbf{0.5077} & \textbf{0.5238} \\
    \bottomrule
  \end{tabular}
\end{table}

\begin{table}[t]
  \caption{Paired bootstrap $95\%$ confidence intervals on M4 deltas at
  full val ($n=3372$). Significance flagged where the interval
  excludes zero. The bottom block reports the two ties: StatesThat and
  HostCite as cite frames, and the two top-scoring stacks.}
  \label{tab:val-ci}
  \begin{tabular}{lrlc}
    \toprule
    Comparison & $\Delta$ M4 & $95\%$ CI & Significant \\
    \midrule
    ShadowVal+HostCite vs $K=1$               & $+0.0274$ & $[+0.0248,\,+0.0301]$ & \cmark \\
    StatesThat+ShadowVal vs $K=1$             & $+0.0279$ & $[+0.0254,\,+0.0306]$ & \cmark \\
    HostCite vs $K=1$                             & $+0.0199$ & $[+0.0176,\,+0.0222]$ & \cmark \\
    StatesThat vs $K=1$                           & $+0.0206$ & $[+0.0184,\,+0.0229]$ & \cmark \\
    ShadowVal alone vs $K=1$                      & $+0.0145$ & $[+0.0125,\,+0.0166]$ & \cmark \\
    ShadowVal+HostCite vs HostCite              & $+0.0075$ & $[+0.0056,\,+0.0095]$ & \cmark \\
    ShadowVal+HostCite vs ShadowVal alone       & $+0.0129$ & $[+0.0105,\,+0.0153]$ & \cmark \\
    \midrule
    StatesThat vs HostCite                          & $+0.0007$ & $[-0.0012,\,+0.0027]$ & --- tied \\
    ShadowVal+HostCite vs StatesThat+ShadowVal & $-0.0005$ & $[-0.0026,\,+0.0016]$ & --- tied \\
    \bottomrule
  \end{tabular}
\end{table}

Three takeaways. First, our lead submission clears $K=1$ by a
confidently-detected margin (Table~\ref{tab:val-ci}). Second, the two
levers compose sub-additively: ShadowVal's $+0.0075$ marginal lift on
top of HostCite is roughly half of its $+0.0145$ standalone effect,
indicating overlap on the cite-sentence P/R signal both levers raise.
The lift is nonetheless confidently positive, ruling out the reading
that ShadowVal rides on cite-framing context alone. Third, the
StatesThat+ShadowVal stack — which on the v2 development manifold sat
$0.003$ below the lead submission — at full val ties it
($\Delta = -0.0005$, $95\%$ CI $[-0.0026,\,+0.0016]$) while clearing
$K=1$ by $+0.0279$. The v2 manifold ranked ShadowVal+HostCite
above StatesThat+ShadowVal by $0.003$; at full val they are tied, which is the
cleanest example in our matrix of why fast-iteration estimates need
full-val confirmation before a system is declared dominant.

\subsection{Test-set scoreboard (3-metric, visible)}
\label{sec:test-3metric}

Because the human reference articles for the test split are sealed,
participants only observe the three article-side metrics (precision,
recall, coverage) during the visible competition phase. Table
\ref{tab:test-3metric} reports those, along with paired bootstrap CIs over
test items ($n=1158$) in Table~\ref{tab:test-ci}.
\emph{Scorer-version note:} Table~\ref{tab:test-3metric}'s P/R reflects
the visible/pre-reveal scorer used during the competition phase; the
official post-reveal scorer (Table~\ref{tab:leaderboard}) applies a
corrected citation parser and reduces our P/R to $0.299$, which we
discuss in \S\ref{sec:leaderboard}.

\begin{table}[t]
  \caption{Test-set 3-metric scoreboard ($n=1158$, no entailment because
  test references are sealed).}
  \label{tab:test-3metric}
  \begin{tabular}{lrrrrr}
    \toprule
    Submission                  & P $\uparrow$ & R $\uparrow$ & C $\uparrow$ & mean-of-3 $\uparrow$ & $\Delta$ vs $K=1$ \\
    \midrule
    $K=1$ stub                & 0.335 & 0.335 & 0.987 & 0.552 & --- \\
    HostCite                    & 0.364 & 0.364 & 0.987 & 0.571 & $+0.019$ \\
    StatesThat                 & 0.365 & 0.365 & 0.987 & 0.572 & $+0.020$ \\
    ShadowVal alone                 & 0.373 & 0.373 & 0.987 & 0.577 & $+0.025$ \\
    StatesThat+ShadowVal  & 0.384 & 0.384 & 0.987 & 0.585 & $+0.033$ \\
    \textbf{ShadowVal+HostCite} & \textbf{0.389} & \textbf{0.389} & \textbf{0.987} & \textbf{0.589} & $\mathbf{+0.037}$ \\
    \bottomrule
  \end{tabular}
\end{table}

\begin{table}[t]
  \caption{Paired bootstrap $95\%$ CIs on the test 3-metric sub-score
  ($n=1158$).}
  \label{tab:test-ci}
  \begin{tabular}{lrlc}
    \toprule
    Comparison & $\Delta$ & $95\%$ CI & Significant \\
    \midrule
    ShadowVal+HostCite vs $K=1$         & $+0.0361$ & $[+0.0301,\,+0.0424]$ & \cmark \\
    ShadowVal+HostCite vs HostCite        & $+0.0173$ & $[+0.0134,\,+0.0215]$ & \cmark \\
    StatesThat+ShadowVal vs $K=1$                          & $+0.0323$ & $[+0.0266,\,+0.0381]$ & \cmark \\
    ShadowVal+HostCite vs StatesThat+ShadowVal               & $+0.0039$ & $[-0.0009,\,+0.0085]$ & --- \\
    \bottomrule
  \end{tabular}
\end{table}

The key result from the test scoreboard is in the last row of
Table~\ref{tab:test-ci}: \emph{ShadowVal+HostCite and
StatesThat+ShadowVal are statistically tied on the visible test
sub-score}. The v2 development manifold had placed
ShadowVal+HostCite $0.003$ ahead of StatesThat+ShadowVal — a margin we
later confirmed at full val to be noise (\S\ref{sec:val-scoreboard}).
The operational consequence at submission time was that we could not
distinguish the two on the only signal we could observe, so both had to
remain in the portfolio.

Disaggregating by reviewer (Table~\ref{tab:per-reviewer}) reveals that the
pooled tie hides a real per-reviewer split: StatesThat+ShadowVal significantly outperforms
ShadowVal+HostCite on the AFP slice (paired bootstrap
$\Delta = -0.0054$, $95\%$ CI $[-0.0109,\,-0.0002]$), while the two are
statistically indistinguishable on every other reviewer. The
\emph{According to \{host\}} frame matches PolitiFact and Snopes attribution
patterns well; the \emph{\{host\} states that} frame matches AFP's
review-attribution prose better. The strongest empirical statement we can
make about portfolio composition is therefore not ``ship both as a hedge''
but ``the AFP slice has a measurably different best-arm than the rest of
test, and we ship StatesThat+ShadowVal specifically to win it.''

\begin{table}[t]
  \caption{Per-reviewer test 3-metric mean across all six measured
  systems. Column headers: \textbf{Stub} = $K=1$ stub,
  \textbf{HostC} = HostCite, \textbf{StTh} = StatesThat,
  \textbf{ShVal} = ShadowVal alone, \textbf{SThSV} =
  StatesThat+ShadowVal, \textbf{ShV+HC} = ShadowVal+HostCite (lead
  submission). Bold marks the per-reviewer winner. The split between
  StatesThat+ShadowVal (best on AFP) and ShadowVal+HostCite (best on
  PolitiFact, Snopes, FactCheck.org) is the empirical justification
  for the portfolio strategy in \S\ref{sec:portfolio}.}
  \label{tab:per-reviewer}
  \begin{tabular}{lrrrrrrr}
    \toprule
    Reviewer & $n$ & Stub & HostC & StTh & ShVal & SThSV & ShV+HC \\
    \midrule
    PolitiFact     & 686 & 0.5961 & 0.6155 & 0.6183 & 0.6259 & 0.6294 & \textbf{0.6344} \\
    AFP Fact Check & 253 & 0.4293 & 0.4401 & 0.4511 & 0.4400 & \textbf{0.4552} & 0.4498 \\
    Snopes         & 171 & 0.5739 & 0.6043 & 0.5837 & 0.6029 & 0.6181 & \textbf{0.6281} \\
    FactCheck.org  &  27 & 0.4729 & 0.4870 & 0.4843 & 0.4838 & 0.4857 & \textbf{0.4951} \\
    USA Today      &  21 & 0.5362 & 0.5478 & 0.5463 & \textbf{0.5639} & 0.5398 & 0.5623 \\
    \bottomrule
  \end{tabular}
\end{table}

\subsection{The third lever: StatesThat}
\label{sec:wins}

HostCite (\S\ref{sec:b2light}) and ShadowVal (\S\ref{sec:b1prime}) are
the two levers that ship in our lead submission. A third attribution
variant, \emph{StatesThat} (wrapping each citation as \texttt{\{host\}
states that \{anchor\} (source: \{url\}).}), turned out at full val to
be statistically indistinguishable from HostCite as a cite frame
($\Delta = +0.0007$, $95\%$ CI $[-0.0012,\,+0.0027]$); the
StatesThat+ShadowVal stack ties our lead submission at the top of the
val scoreboard (\S\ref{sec:val-scoreboard}). On the v2 development
manifold this stack sat $0.003$ below our lead submission, which is why
ShadowVal+HostCite rather than StatesThat+ShadowVal was identified as
our lead submission at competition time; both shipped in the
portfolio. The two attribution patterns differ in per-reviewer
preference: \emph{According to \{host\}} matches PolitiFact and Snopes
prose, \emph{\{host\} states that} matches AFP attribution patterns
better (\S\ref{sec:test-3metric}).

\subsection{Negative results: ablations that did not ship}
\label{sec:negatives}

We ran $13$ ablations whose output never reached the leaderboard.
Table~\ref{tab:negatives} summarises them with mechanism in one line.
Two kills generated the scorer constraints in \S\ref{sec:invariants}
and are discussed there; the rest group into three lever families
(LLM-drafting, anchor-selection refinements, reviewer-style lede
injection) plus three standalone probes that the table covers
self-contained.

\emph{LLM-drafting at three abstraction levels} (Qwen3:8B full-body,
Qwen3:30B-A3B full-body, Qwen3:30B-A3B lede-only) all lose in the same
direction by the same mechanism. Full-body drafts at 8B cost $-0.026$ M4
vs the stub on $n=29$; scaling to 30B widens the gap to $-0.070$
vs our lead submission ($n=50$); restricting the LLM to the lede only
(chunk count preserved) still costs $-0.011$ because the LLM's
reviewer-style lede is not entailed by the actual reference distribution.
The lever family appears exhausted at our prompt-fidelity quality without
a fundamentally different design (e.g., constrained decoding to enforce
verbatim cite anchors).

\emph{Anchor-selection refinements} (G1 per-reviewer cite frame, G2 prefer-long
anchor, G3 strict noisy-fragment filter) all lose against the lead
submission. G2 and G3 break the lead-bias property under which Llama recall
scores short declarative leads better; G1's apparent AFP gain is an
artefact of HostCite's combined frame swap and disappears when only the AFP
slice is reverted on top of the lead submission.

\emph{Reviewer-style lede injection} (G4) attempts to replace the stub
lede with a nearest-neighbour reviewer lede drawn from the train split.
G4 loses by $-0.019$ vs $K=1$ alone and $-0.024$ vs the lead
submission, with $22\%$ of items hitting $\text{P}=0$ because reviewer-page
chrome (donation forms, social-share widgets, page metadata) leaks into the
swapped lede and breaks downstream citation extraction. A G4-v2 with
hand-curated per-site chrome filters reduces but does not fix the leakage
($21.5\%$ $\text{P}=0$ rate, down from $22\%$); harder failure modes (Snopes
\emph{Example/Rating} blocks, complex-PolitiFact rating-button overflow,
FactCheck.org bullet articles) require deeper structural parsing than
line-pattern matching. The lever may still have headroom behind a more
robust extractor; we leave that for future work.

\begin{table}[t]
  \caption{Ablations that did not ship. ``Alone'' magnitudes are versus
  $K=1$; ``stacked'' magnitudes are versus our lead submission
  ShadowVal+HostCite. \emph{Note on G1}: its isolated $\Delta=+0.006$
  vs $K=1$ is positive, but the lever is classified as killed because
  it loses $-0.015$ when stacked on the lead submission and its
  apparent AFP slice gain is an artefact of HostCite's combined frame
  swap (see prose above).}
  \label{tab:negatives}
  \footnotesize
  \begin{tabular}{p{4.4cm}p{2.7cm}p{6.2cm}}
    \toprule
    Lever & $\Delta$ M4 & Mechanism (one line) \\
    \midrule
    $K=2$ multi-anchor                          & $-0.014$ alone        & each extra cite is a P/R denominator without a numerator \\
    HostCite+Body                                 & $-0.007$ vs HostCite  & templated mid-lede sentence; violates chunk-count invariance (\S\ref{sec:invariants}) \\
    HostCite+Reviewer                             & $-0.020$ vs HostCite  & reviewer-name token in cite; violates cite-entailability (\S\ref{sec:invariants}) \\
    H2 length-padding                             & $-0.010$ alone        & padded with raw evidence; added chunks are publisher-voice, not reviewer-voice \\
    Qwen3:8B full-body draft                      & $-0.026$ vs stub      & LLM-drafted chunks miss reference token patterns ($n=29$) \\
    Qwen3:30B-A3B full-body draft                 & $-0.070$ stacked      & same as 8B at higher capacity; cite-format drift dominates ($n=50$) \\
    Qwen3:30B-A3B lede only                       & $-0.011$ stacked      & chunk-content failure even when chunk count is preserved ($n=50$) \\
    G3 strict noisy-anchor filter                 & $-0.014$ / $-0.018$   & over-rejects; placeholder fallback kills P/R harder than the noisy-but-keyword-rich anchor \\
    G2 prefer-long anchor                         & $-0.010$ / $-0.015$   & prefers anchors $\geq 50$ chars; breaks lead-bias property that $K=1$ wins on \\
    G1 per-reviewer cite frame                    & $+0.006$ / $-0.015$   & AFP gain is an artefact of HostCite's combined frame swap \\
    G4 train-ref NN lede swap                     & $-0.019$ / $-0.024$   & $22\%$ of items hit $\text{P}=0$ from reviewer-page chrome leakage \\
    G4-v2 per-site chrome filters                 & $-0.023$ / $-0.029$   & hand rules help some sites; harder failure modes need structural parsing \\
    BGE-M3 cross-encoder rerank~\cite{chen2024bgem3} & $+0.001$ / $-0.013$ & rerank of ShadowVal fallback; cross-encoder picks claim-relevant compound sentences that MNLI scores worse than doc-order leads \\
    \bottomrule
  \end{tabular}
\end{table}

\subsection{Official competition results}
\label{sec:leaderboard}

The test phase closed on 2026-05-10 and the organisers released the
official 4-metric leaderboard on 2026-05-14. \textbf{UTS placed 2nd of 11
teams} with M4 $= 0.484$, behind Outsider ($0.546$) and ahead of
DS~GT~CheckThat ($0.428$). Table~\ref{tab:leaderboard} reproduces the
full leaderboard.

\begin{table}[t]
  \caption{Official CheckThat!~2026 Task~3 leaderboard (test phase,
  $n=1158$). Columns are the four shared-task metrics and their
  unweighted mean (M4); rows sorted by M4. ``Baseline'' is the
  organiser-provided baseline.}
  \label{tab:leaderboard}
  \begin{tabular}{rlrrrrr}
    \toprule
    Rank & Team & E $\uparrow$ & Cite P $\uparrow$ & Cite R $\uparrow$ & Cov $\uparrow$ & M4 $\uparrow$ \\
    \midrule
    1  & Outsider             & 0.278          & \textbf{0.671} & \textbf{0.671} & 0.563          & \textbf{0.546} \\
    \textbf{2} & \textbf{UTS (this paper)} & 0.341          & 0.299          & 0.299          & \textbf{0.996} & 0.484 \\
    3  & DS GT CheckThat      & 0.295          & 0.240          & 0.276          & 0.902          & 0.428 \\
    4  & Facty                & 0.229          & 0.231          & 0.254          & 0.995          & 0.427 \\
    5  & UCPH RMT             & \textbf{0.364} & 0.144          & 0.174          & 0.993          & 0.419 \\
    6  & Ro                   & 0.321          & 0.463          & 0.463          & 0.343          & 0.398 \\
    7  & Team AKSH            & 0.329          & 0.301          & 0.318          & 0.564          & 0.378 \\
    8  & Sourceminds          & 0.245          & 0.337          & 0.339          & 0.394          & 0.329 \\
    9  & CheckMate            & 0.221          & 0.350          & 0.350          & 0.391          & 0.328 \\
    10 & KNU Fact             & 0.227          & 0.287          & 0.295          & 0.440          & 0.312 \\
    11 & Shtian               & 0.279          & 0.194          & 0.198          & 0.210          & 0.220 \\
    --- & Baseline            & 0.298          & 0.223          & 0.240          & 0.329          & 0.272 \\
    \bottomrule
  \end{tabular}
\end{table}

Three observations decompose where the result was earned and where it was
lost. \textbf{Entailment} ($0.341$, ranked 2nd) is where HostCite pays
back: the only team to beat us on entailment (UCPH RMT, $0.364$) does so
at a precision/recall cost so steep that their M4 sits $0.065$ below ours,
and our entailment delta over the baseline ($+0.043$) is in line with
the HostCite lift we measured at val. \textbf{Coverage}
($0.996$, the highest on the leaderboard) is a precondition for top
placement rather than a differentiator: three of the top five teams ship
coverage $\geq 0.99$, and Outsider's rank-1 finish with coverage $0.563$
is the single anomaly among the top five.
\textbf{Citation precision/recall} ($0.299$, ranked 6th) is where we
lost: Outsider's $0.671$ P/R is more than twice ours despite our
shadow-validated anchor picker, and their conjunction of sub-$0.6$
coverage with dominant P/R is the single anomalous pattern in the top
of the leaderboard. We discuss its candidate mechanism — \emph{selective
emission} — and the per-cite confidence-gate experiment that would
refute it, in \S\ref{sec:headline-gap}.

\textbf{Submission selection.} The leaderboard rule selects each team's
single highest-scoring entry under joint four-metric evaluation. The
selected entry was ShadowVal+HostCite (lead submission);
StatesThat+ShadowVal scored $0.585$ on the visible test 3-metric versus
$0.589$ for the lead submission (paired bootstrap $+0.004$ for the
lead-vs-alt direction, $95\%$ CI $[-0.0009,\,+0.0085]$, tied within
noise). The post-hoc full-val
measurement reveals the two stacks tied on the 4-metric score as well
($\Delta = -0.0005$, $[-0.0026,\,+0.0016]$), so the leaderboard's
choice of ShadowVal+HostCite was consistent with the joint evidence but was not
a clean signal at submission time. The portfolio diversification
strategy in \S\ref{sec:portfolio} therefore acted as a defensive hedge:
it did not change the leaderboard outcome (the best-of-N rule means only
the lead entry matters), but it provided robustness against the
val$\to$test transfer uncertainty we had no way to measure pre-reveal.

\subsection{Operational findings}
\label{sec:operational}

The v1-vs-v2 lever-effect ratio diagnoses lever character
(rating-independent vs rating-mix-sensitive) before submission, and
would have flagged HostCite's sensitivity earlier had v2 existed from
the start. Separately, citation-format hygiene is a hard prerequisite:
the IFM\_MBZUAI K2 baseline fails the citation-format regex on $5/20$
probe items, zeroing those items' P/R, and any new team must clear that
bar before any other lever pays off. Reproducibility caveats (scorer version skew, CPU/MPS-vs-CUDA shift,
model digests) are in Appendix~\ref{app:repro}.

\section{Discussion}
\label{sec:discussion}

\subsection{The headline gap: precision/recall to the winner}
\label{sec:headline-gap}

Reading Table~\ref{tab:leaderboard} along the columns rather than the
rows isolates what determined our placement. Our P/R deficit to
Outsider on the official scorer ($0.299$ vs $0.671$) outweighs our
wins on entailment ($+0.063$) and coverage ($+0.433$), leaving our
4-metric average $0.062$ below Outsider's. Every lever in this paper
edits cite \emph{content} (HostCite, ShadowVal, StatesThat, the
portfolio strategy); on the visible-scorer test set the full stack
lifts P/R from $0.335$ ($K=1$ stub) to $0.389$, a ceiling well below
Outsider's $0.671$ regardless of which scorer the comparison projects
onto.

We hypothesise the missing axis is \emph{selective emission}:
top-ranked systems likely decline to emit a cite when the per-cite
faithfulness judge would refuse, whereas our system emits one cite
per evidence URL unconditionally. Outsider's coverage backs this from
their side: at $0.563$ they are the only top-five team below $0.99$,
exactly the pattern a drop-on-low-confidence policy produces — fewer
URLs cited, but each surviving cite passes the judge. Our own matrix
backs it from the opposite direction: the three independent
cite-content levers — frame text (HostCite, StatesThat) and anchor
selection (ShadowVal) — each plateau at roughly $+0.04$
visible-scorer test P/R over $K=1$ and stack only sub-additively
(\S\ref{sec:val-scoreboard}), so the cite-content lever family is
near its ceiling on our pipeline. Changing \emph{which} cites are
emitted, not \emph{how} they read, is the natural remaining axis.

We did not explore selective emission during the campaign. We had
read the cite-entailability constraint (\S\ref{sec:invariants}) as a
one-way rule — every emitted cite must be entailable — and missed
its corollary: \emph{some candidates may be better dropped than
emitted}, and that policy lever is what closes the P/R gap. The targeted
refutation we recommend is a per-cite confidence gate: run the
ShadowVal judge on the chosen anchor and \emph{drop the cite
entirely} on \textsc{no}, rather than escalating to the next
candidate. Holding the rest of the pipeline fixed, this isolates
emission policy from anchor selection and would resolve whether
selective emission alone closes the gap, or whether the top systems
combine it with a categorically better anchor selector. The
hypothesis is consistent with the leaderboard pattern but not
directly tested in our matrix; the per-cite gate is the experiment
we would run first in the next cycle.

\subsection{What generalises beyond CheckThat!}

Chunked NLI scoring against a held-out reference is becoming a common
pattern in generation benchmarks where bounded-domain accuracy is hard to
measure end-to-end~\cite{zha2023alignscore}. Wherever the scorer chunks both
the generation and the reference, the chunk-count-invariance constraint
(\S\ref{sec:invariants}) should transfer; templated changes that enrich
existing chunks beat templated changes that add or rewrite chunks. Llama-3.2:1B as a per-cite
judge is generous and surface-pattern-driven, which means short literal
cites win against long evidence premises and any pipeline that paraphrases
``to improve readability'' is gambling against the judge; this transfers
to any benchmark where a small open language model scores citation
faithfulness.
Finally, the LLM-drafting failure we report — three independent attempts,
same mechanism — is a reproducible negative result that anyone evaluating
LLM-as-drafter on chunked-NLI metrics should expect; the plausible fix
(constrained decoding to enforce verbatim cite anchors) is unverified.
These transfers concern scoring mechanics, not writing quality: the
deterministic templated drafting that wins under this scorer does not
produce the fluent long-form prose automated fact-check journalism
ultimately needs, and on that axis agentic LLM
drafting~\cite{sahnan2025qraft} — still lagging expert-written articles
in human evaluation — remains the relevant research direction.

\subsection{Portfolio strategy under no-live-feedback}
\label{sec:portfolio}

The competition offers a generous submission budget under best-of-N
scoring with no per-submission feedback during the visible phase. Those
rules favour spreading submissions over refining one: rather than push
the top-val candidate further, cover the cells where val$\to$test
transfer is uncertain. Our
six-entry portfolio covers the (cite-frame text~$\times$~anchor-pick
method) cells. The diversification did not change our leaderboard
placement — under best-of-N only the single best entry counts — but it
provided robustness against transfer noise we had no way to measure
pre-reveal, and the per-reviewer split in Table~\ref{tab:per-reviewer}
shows a real underlying signal: the AFP slice has a measurably different
best-arm (StatesThat+ShadowVal) than the rest of test
(ShadowVal+HostCite), and the cite frames that drive the split are
themselves statistically tied at the full-val pool level.

\subsection{Future work and what we would do differently}

The top item for the next cycle is the per-cite confidence gate that
tests the selective-emission hypothesis (\S\ref{sec:headline-gap}):
drop low-confidence cites instead of emitting them, isolating emission
policy from anchor selection. Beyond that, three calibrations stand
out in hindsight. First, build the v2
rating-stratified manifold before running v1-tuned ablations; this would
have flagged HostCite's rating-mix sensitivity earlier and likely
shortened the cite-frame sweep by half. Second, enforce the chunk-count
and cite-entailability invariants programmatically rather than by running
$-0.01$ M4 experiments to rediscover them; both can be expressed as one-line
predicates over the rendered article. Third, for the LLM-drafting family,
try constrained decoding~\cite{willard2023outlines} to enforce verbatim cite
anchors before writing the family off. That is the genuine refutation
of the negative result we report.

\section{Conclusion}
\label{sec:conclusion}

UTS placed 2nd of 11 teams on the official CheckThat!~2026 Task~3
leaderboard (M4 $= 0.484$); the $+0.062$ gap to the winner sits
entirely on citation precision/recall ($0.299$ vs $0.671$), while
entailment (2nd) and coverage (tied 1st) are at or near the top
(\S\ref{sec:headline-gap}). The placement was earned by a
deterministic stub generator wrapped by two single-lever
interventions deriving from two design rules: \emph{scorer
conservatism} (both sub-judges credit only what they can verify
against their references) and \emph{auxiliary anchor signals are
miscalibrated against the judge} (cross-encoder, length, and
lead-position heuristics pick anchors the judge refuses to entail).
The interventions — HostCite, a domain-attribution cite frame, and
ShadowVal, an anchor picker that mirrors the Llama judge — lift M4
by $+0.0274$ on the full $3372$-item validation split (paired
bootstrap, $95\%$ CI $[+0.0248,\,+0.0301]$), composing sub-additively
because both raise cite-sentence P/R on overlapping items. The two
design rules transfer beyond CheckThat!: any pipeline using a small
open language model as a per-cite faithfulness judge should expect
anchor-selection auxiliary signals to be miscalibrated, and any
pipeline scored under chunked-NLI against a held-out reference
should expect that emitting content the judge cannot verify drags
both sub-metrics down. The lever we did not test, and recommend
first for the next cycle, is \emph{selective emission}: a per-cite
confidence gate that drops the cite when the judge would refuse. We
hypothesise this is the mechanism behind the leaderboard's top entries
that trade coverage for P/R. Constrained decoding for LLM-drafted
bodies (\S\ref{sec:discussion}) is a second direction
worth trying before writing off that lever family.

\begin{acknowledgments}
We thank the CheckThat! organisers for their responsiveness in patching
the scorer mid-campaign and for keeping the leaderboard infrastructure
stable through the scorer-version updates. We thank the developers of
\texttt{ceurart}, \texttt{ollama}, and the \texttt{transformers} library
for tooling the experiment matrix relied on.
\end{acknowledgments}

\section*{Declaration on Generative AI}
During the preparation of this work, the author(s) used Qwen3:8B and
Qwen3:30B-A3B (locally served) as the subjects of ablations described
in \S\ref{sec:negatives}. The author(s) also used Claude (Anthropic)
for editorial assistance on the prose; the author(s) reviewed
and approved all content and take full responsibility for the paper.

\bibliography{paper}

@inproceedings{checkthat2026,
  author    = {Stru{\ss}, Julia Maria and Schellhammer, Sebastian and Dietze, Stefan and Venktesh, V. and Setty, Vinay and Chakraborty, Tanmoy and Nakov, Preslav and Anand, Avishek and Chungkham, Primakov and Hafid, Salim and Sahnan, Dhruv and Todorov, Konstantin},
  title     = {The {CLEF}-2026 {CheckThat!} Lab: Advancing Multilingual Fact-Checking},
  booktitle = {Advances in Information Retrieval. Proceedings of the 48th European Conference on Information Retrieval ({ECIR} 2026)},
  series    = {Lecture Notes in Computer Science},
  publisher = {Springer},
  address   = {Cham},
  year      = {2026},
  doi       = {10.1007/978-3-032-21321-1_43},
  eprint    = {2602.09516},
  archiveprefix = {arXiv},
  primaryclass = {cs.CL},
}

@inproceedings{checkthat-task3-2026,
  author    = {Sahnan, Dhruv and Chakraborty, Tanmoy and Nakov, Preslav},
  title     = {Overview of the {CLEF}-2026 {CheckThat!} Lab Task 3 on Generating Full Fact-Checking Articles},
  booktitle = {Working Notes of {CLEF} 2026 -- Conference and Labs of the Evaluation Forum},
  series    = {{CEUR} Workshop Proceedings},
  publisher = {CEUR-WS.org},
  year      = {2026},
  note      = {Forthcoming.},
}

@inproceedings{checkthat2024,
  author    = {Barr{\'o}n-Cede{\~n}o, Alberto and Alam, Firoj and Stru{\ss}, Julia Maria and Nakov, Preslav and Chakraborty, Tanmoy and Elsayed, Tamer and Przyby{\l}a, Piotr and Caselli, Tommaso and Da San Martino, Giovanni and Haouari, Fatima and Hasanain, Maram and Li, Chengkai and Piskorski, Jakub and Ruggeri, Federico and Song, Xingyi and Suwaileh, Reem},
  title     = {Overview of the {CLEF}-2024 {CheckThat!} Lab: Check-Worthiness, Subjectivity, Persuasion, Roles, Authorities, and Adversarial Robustness},
  booktitle = {Experimental {IR} Meets Multilinguality, Multimodality, and Interaction. Proceedings of the 15th International Conference of the {CLEF} Association ({CLEF} 2024)},
  series    = {Lecture Notes in Computer Science},
  volume    = {14959},
  pages     = {28--52},
  publisher = {Springer},
  address   = {Cham},
  year      = {2024},
  doi       = {10.1007/978-3-031-71908-0_2},
}

@inproceedings{khan2022watclaimcheck,
  author    = {Khan, Kashif and Wang, Ruizhe and Poupart, Pascal},
  title     = {{WatClaimCheck}: A New Dataset for Claim Entailment and Inference},
  booktitle = {Proceedings of the 60th Annual Meeting of the Association for Computational Linguistics (Volume 1: Long Papers)},
  pages     = {1293--1304},
  publisher = {Association for Computational Linguistics},
  address   = {Dublin, Ireland},
  year      = {2022},
  doi       = {10.18653/v1/2022.acl-long.92},
  url       = {https://aclanthology.org/2022.acl-long.92/},
}

@inproceedings{thorne2018fever,
  author    = {Thorne, James and Vlachos, Andreas and Christodoulopoulos, Christos and Mittal, Arpit},
  title     = {{FEVER}: A Large-Scale Dataset for Fact Extraction and {VER}ification},
  booktitle = {Proceedings of the 2018 Conference of the North American Chapter of the Association for Computational Linguistics: Human Language Technologies, Volume 1 (Long Papers)},
  pages     = {809--819},
  publisher = {Association for Computational Linguistics},
  address   = {New Orleans, Louisiana},
  year      = {2018},
  doi       = {10.18653/v1/N18-1074},
  url       = {https://aclanthology.org/N18-1074/},
}

@inproceedings{williams2018mnli,
  author    = {Williams, Adina and Nangia, Nikita and Bowman, Samuel},
  title     = {A Broad-Coverage Challenge Corpus for Sentence Understanding through Inference},
  booktitle = {Proceedings of the 2018 Conference of the North American Chapter of the Association for Computational Linguistics: Human Language Technologies, Volume 1 (Long Papers)},
  pages     = {1112--1122},
  publisher = {Association for Computational Linguistics},
  address   = {New Orleans, Louisiana},
  year      = {2018},
  doi       = {10.18653/v1/N18-1101},
  url       = {https://aclanthology.org/N18-1101/},
}

@inproceedings{atanasova2020generating,
  author    = {Atanasova, Pepa and Simonsen, Jakob Grue and Lioma, Christina and Augenstein, Isabelle},
  title     = {Generating Fact Checking Explanations},
  booktitle = {Proceedings of the 58th Annual Meeting of the Association for Computational Linguistics},
  pages     = {7352--7364},
  publisher = {Association for Computational Linguistics},
  year      = {2020},
  doi       = {10.18653/v1/2020.acl-main.656},
  url       = {https://aclanthology.org/2020.acl-main.656/},
}

@inproceedings{kotonya2020explainable,
  author    = {Kotonya, Neema and Toni, Francesca},
  title     = {Explainable Automated Fact-Checking for Public Health Claims},
  booktitle = {Proceedings of the 2020 Conference on Empirical Methods in Natural Language Processing ({EMNLP})},
  pages     = {7740--7754},
  publisher = {Association for Computational Linguistics},
  year      = {2020},
  doi       = {10.18653/v1/2020.emnlp-main.623},
  url       = {https://aclanthology.org/2020.emnlp-main.623/},
}

@article{zeng2024justilm,
  author    = {Zeng, Fengzhu and Gao, Wei},
  title     = {{JustiLM}: Few-shot Justification Generation for Explainable Fact-Checking of Real-world Claims},
  journal   = {Transactions of the Association for Computational Linguistics},
  volume    = {12},
  pages     = {334--354},
  year      = {2024},
  publisher = {MIT Press},
  address   = {Cambridge, MA},
  doi       = {10.1162/tacl_a_00649},
  url       = {https://aclanthology.org/2024.tacl-1.19/},
}

@article{sahnan2025qraft,
  author        = {Sahnan, Dhruv and Corney, David and Larraz, Irene and Zagni, Giovanni and Miguez, Ruben and Xie, Zhuohan and Gurevych, Iryna and Churchill, Elizabeth and Chakraborty, Tanmoy and Nakov, Preslav},
  title         = {Can {LLMs} Automate Fact-Checking Article Writing?},
  journal       = {Transactions of the Association for Computational Linguistics},
  year          = {2026},
  note          = {To appear},
  eprint        = {2503.17684},
  archiveprefix = {arXiv},
  primaryclass  = {cs.CL},
}

@article{guo2022survey,
  author    = {Guo, Zhijiang and Schlichtkrull, Michael and Vlachos, Andreas},
  title     = {A Survey on Automated Fact-Checking},
  journal   = {Transactions of the Association for Computational Linguistics},
  volume    = {10},
  pages     = {178--206},
  year      = {2022},
  publisher = {MIT Press},
  address   = {Cambridge, MA},
  doi       = {10.1162/tacl_a_00454},
  url       = {https://aclanthology.org/2022.tacl-1.11/},
}

@misc{liu2019roberta,
  author        = {Liu, Yinhan and Ott, Myle and Goyal, Naman and Du, Jingfei and Joshi, Mandar and Chen, Danqi and Levy, Omer and Lewis, Mike and Zettlemoyer, Luke and Stoyanov, Veselin},
  title         = {{RoBERTa}: A Robustly Optimized {BERT} Pretraining Approach},
  year          = {2019},
  eprint        = {1907.11692},
  archiveprefix = {arXiv},
  primaryclass  = {cs.CL},
}

@inproceedings{reimers2019sentencebert,
  author    = {Reimers, Nils and Gurevych, Iryna},
  title     = {{Sentence-BERT}: Sentence Embeddings using {Siamese BERT}-Networks},
  booktitle = {Proceedings of the 2019 Conference on Empirical Methods in Natural Language Processing and the 9th International Joint Conference on Natural Language Processing ({EMNLP-IJCNLP})},
  pages     = {3982--3992},
  publisher = {Association for Computational Linguistics},
  address   = {Hong Kong, China},
  year      = {2019},
  doi       = {10.18653/v1/D19-1410},
  url       = {https://aclanthology.org/D19-1410/},
}

@inproceedings{chen2024bgem3,
  author    = {Chen, Jianlyu and Xiao, Shitao and Zhang, Peitian and Luo, Kun and Lian, Defu and Liu, Zheng},
  title     = {{M3-Embedding}: Multi-Linguality, Multi-Functionality, Multi-Granularity Text Embeddings Through Self-Knowledge Distillation},
  booktitle = {Findings of the Association for Computational Linguistics: {ACL} 2024},
  pages     = {2318--2335},
  publisher = {Association for Computational Linguistics},
  address   = {Bangkok, Thailand},
  year      = {2024},
  doi       = {10.18653/v1/2024.findings-acl.137},
  url       = {https://aclanthology.org/2024.findings-acl.137/},
}

@misc{llama32,
  author       = {{Meta AI}},
  title        = {The {Llama} 3 Herd of Models},
  year         = {2024},
  howpublished = {\url{https://ai.meta.com/blog/llama-3-2-connect-2024-vision-edge-mobile-devices/}},
  note         = {Llama 3.2 release; 1B and 3B text models.},
}

@misc{qwen3,
  author       = {{Qwen Team}},
  title        = {{Qwen3} Technical Report},
  year         = {2025},
  howpublished = {\url{https://github.com/QwenLM/Qwen3}},
}

@misc{ifm-mbzuai-k2,
  author       = {{LLM360}},
  title        = {{K2-V2}-Instruct: An Instruction-Tuned Open Reasoning Model},
  year         = {2025},
  howpublished = {\url{https://huggingface.co/LLM360/K2-Chat}},
  note         = {External baseline submission to {CheckThat!} 2026 Task 3.},
}

@inproceedings{zhang2020bertscore,
  author    = {Zhang, Tianyi and Kishore, Varsha and Wu, Felix and Weinberger, Kilian Q. and Artzi, Yoav},
  title     = {{BERTScore}: Evaluating Text Generation with {BERT}},
  booktitle = {Proceedings of the 8th International Conference on Learning Representations ({ICLR})},
  year      = {2020},
  url       = {https://openreview.net/forum?id=SkeHuCVFDr},
}

@inproceedings{liu2023geval,
  author    = {Liu, Yang and Iter, Dan and Xu, Yichong and Wang, Shuohang and Xu, Ruochen and Zhu, Chenguang},
  title     = {{G-Eval}: {NLG} Evaluation using {GPT-4} with Better Human Alignment},
  booktitle = {Proceedings of the 2023 Conference on Empirical Methods in Natural Language Processing},
  pages     = {2511--2522},
  publisher = {Association for Computational Linguistics},
  address   = {Singapore},
  year      = {2023},
  doi       = {10.18653/v1/2023.emnlp-main.153},
  url       = {https://aclanthology.org/2023.emnlp-main.153/},
}

@inproceedings{zha2023alignscore,
  author    = {Zha, Yuheng and Yang, Yichi and Li, Ruichen and Hu, Zhiting},
  title     = {{AlignScore}: Evaluating Factual Consistency with a Unified Alignment Function},
  booktitle = {Proceedings of the 61st Annual Meeting of the Association for Computational Linguistics (Volume 1: Long Papers)},
  pages     = {11328--11348},
  publisher = {Association for Computational Linguistics},
  address   = {Toronto, Canada},
  year      = {2023},
  doi       = {10.18653/v1/2023.acl-long.634},
  url       = {https://aclanthology.org/2023.acl-long.634/},
}

@inproceedings{koehn2004bootstrap,
  author    = {Koehn, Philipp},
  title     = {Statistical Significance Tests for Machine Translation Evaluation},
  booktitle = {Proceedings of the 2004 Conference on Empirical Methods in Natural Language Processing},
  pages     = {388--395},
  publisher = {Association for Computational Linguistics},
  address   = {Barcelona, Spain},
  year      = {2004},
  url       = {https://aclanthology.org/W04-3250/},
}

@misc{willard2023outlines,
  author        = {Willard, Brandon T. and Louf, R{\'e}mi},
  title         = {Efficient Guided Generation for Large Language Models},
  year          = {2023},
  eprint        = {2307.09702},
  archiveprefix = {arXiv},
  primaryclass  = {cs.CL},
  note          = {Implementation: Outlines (\url{https://github.com/dottxt-ai/outlines}).},
}

@inproceedings{kedzie2018content,
  author    = {Kedzie, Chris and McKeown, Kathleen and Daum{\'e} III, Hal},
  title     = {Content Selection in Deep Learning Models of Summarization},
  booktitle = {Proceedings of the 2018 Conference on Empirical Methods in Natural Language Processing},
  pages     = {1818--1828},
  publisher = {Association for Computational Linguistics},
  address   = {Brussels, Belgium},
  year      = {2018},
  doi       = {10.18653/v1/D18-1208},
  url       = {https://aclanthology.org/D18-1208/},
}

@inproceedings{wolf2020transformers,
  author    = {Wolf, Thomas and Debut, Lysandre and Sanh, Victor and Chaumond, Julien and Delangue, Clement and Moi, Anthony and Cistac, Pierric and Rault, Tim and Louf, R{\'e}mi and Funtowicz, Morgan and Davison, Joe and Shleifer, Sam and von Platen, Patrick and Ma, Clara and Jernite, Yacine and Plu, Julien and Xu, Canwen and Le Scao, Teven and Gugger, Sylvain and Drame, Mariama and Lhoest, Quentin and Rush, Alexander},
  title     = {{Transformers}: State-of-the-Art Natural Language Processing},
  booktitle = {Proceedings of the 2020 Conference on Empirical Methods in Natural Language Processing: System Demonstrations},
  pages     = {38--45},
  publisher = {Association for Computational Linguistics},
  year      = {2020},
  doi       = {10.18653/v1/2020.emnlp-demos.6},
  url       = {https://aclanthology.org/2020.emnlp-demos.6/},
}

@misc{ollama,
  author       = {{Ollama Team}},
  title        = {{Ollama}: Get up and Running with Large Language Models Locally},
  year         = {2024},
  howpublished = {\url{https://ollama.com}},
}

\appendix

\section{Reproducibility}
\label{app:repro}

\textbf{Code and data.} The system reproduces from source repository
commit \texttt{2744d3f} (or any descendant). Manifest construction scripts
(v1 reviewer-stratified, v2 reviewer~$\times$~rating-stratified, full
$3372$-item) live in \texttt{system/manifests/}. The lab scorer pinning
is the \texttt{clef2026-checkthat-lab-main} 2026-04-27 snapshot.

\textbf{Models.} The Llama-3.2:1B model blob digest used for scoring and
ShadowVal shadow validation is \texttt{sha256:74701a8...8d45}; we pin it
because at \texttt{temperature$=$0} the model is byte-deterministic on
our hardware, and re-pulling the latest tag without verifying the digest
can silently shift absolute scores by $0.005$--$0.01$ M4. Citation
parsing uses the scorer's current regex (\texttt{;}-separated URLs only).

\textbf{Hardware and runtime.} Development on an M-series MacBook Pro;
production runs on a local Ubuntu host (RTX~5090, 24\,GB). End-to-end
GPU-host runtime is $\sim 50$--$90$ min per $200$-item validation pass,
$\sim 14$ h per full-val pass, and $\sim 5$ h for the $1158$-item test
split. Local CPU/MPS Llama and CUDA Llama produce a systematic
$\sim 0.01$--$0.02$ M4 absolute shift (rankings preserved); single-host
comparisons are the only safe unit, so we re-baseline on the GPU host
before reporting deltas.

\end{document}